\documentstyle[NATO,epsf]{CRCKAPB}

\begin{opening}
\title{Progress in modeling magnetic white dwarfs}

\author{S.~Jordan}
\institute{Institut f\"ur Astronomie und Astrophysik, Universit\"at T\"ubingen,
Sand 1, D-72076 T\"ubingen, Germany }

\end{opening}

\begin{document}


\section{Introduction}
The analysis of magnetic white dwarfs has made considerable progress since the
last white dwarf workshop: 
1.  It could be shown that neither the magnetic white dwarf 
Grw~$+70^\circ 8247$ 
\cite{JorFri01} nor LP~790$-$29 (\citeauthor{JorFri02} \citeyear{JorFri02},
\citeauthor{Beuer02} \citeyear{Beuer02})
 are
fast rotators ($P<10$ min) but are rotating slowly ($P>20$ years).
2. 
Automatic fitting procedures have been developed which can
be used to derive the geometries of complex magnetic fields for rotating
white dwarfs (see Euchner et al. 2002, Euchner et al., these proceedings,
G\"ansicke et al, these proceedings). 
3.  First models with consistent energy values and
oscillator strengths for neutral helium in strong magnetic fields have
been calculated and can be compared to observed spectra and polarization
data.

Grw~$+70^\circ 8247$ the first white dwarf in which a magnetic field
has been detected, turned out to be a long standing challenge for
the modeling, since
the polarization could not be fitted, while the flux spectrum could 
be well reproduced by centered dipole models. Now, a possible solution
could be found.
First models with consistent energy values and
oscillator strengths for neutral helium in strong magnetic fields have
been calculated and can be compared to observed spectra and polarization
data. Such calculation were applied to the DAP GD~229. 

\section{Analysis of Grw~$+70^\circ 8247$}
Fits to the flux spectrum of Grw~$+70^\circ 8247$ have been published by
\citeauthor{WF88} \shortcite{WF88}, \citeauthor{JorPhD} \shortcite{JorPhD},
 \citeauthor{JorDart} \shortcite{JorDart}, and \citeauthor{Jor92} \shortcite{Jor92}.
The result was a  pure dipole models with a  polar
field strength of about 320\,MG. However, one mystery of this well
 studied object still remained: the detailed
wavelength dependence of the circular and linear polarization predicted by the
dipole model strongly deviated from the observation.

Since the fitting procedure has very much improved since than, 
I have repeated the analysis  with a
method developed by Jordan and Rahn. The code is  similar to the one
described by  \cite{euchneretal02-1} but uses a genetic algorithm instead
of an evolutionary strategy.

The best fit of the spectrum to a pure magnetic dipole model resulted in
a polar field strength of 347\,MG and an angle $i=56^\circ$ between
the dipole axis and the observer. However, the model cannot 
fit the observed  polarization 
(see. Fig.\,1). 
The situation does not change if one allows for offsets of the dipole
relative to the center of the star. The fitting of the spectrum resulted
in  dipole offsets smaller than 0.005 stellar radii.

\begin{figure}[t]
\epsfxsize=0.35\textwidth
\epsfbox[51.6 42.0 350 652.5]{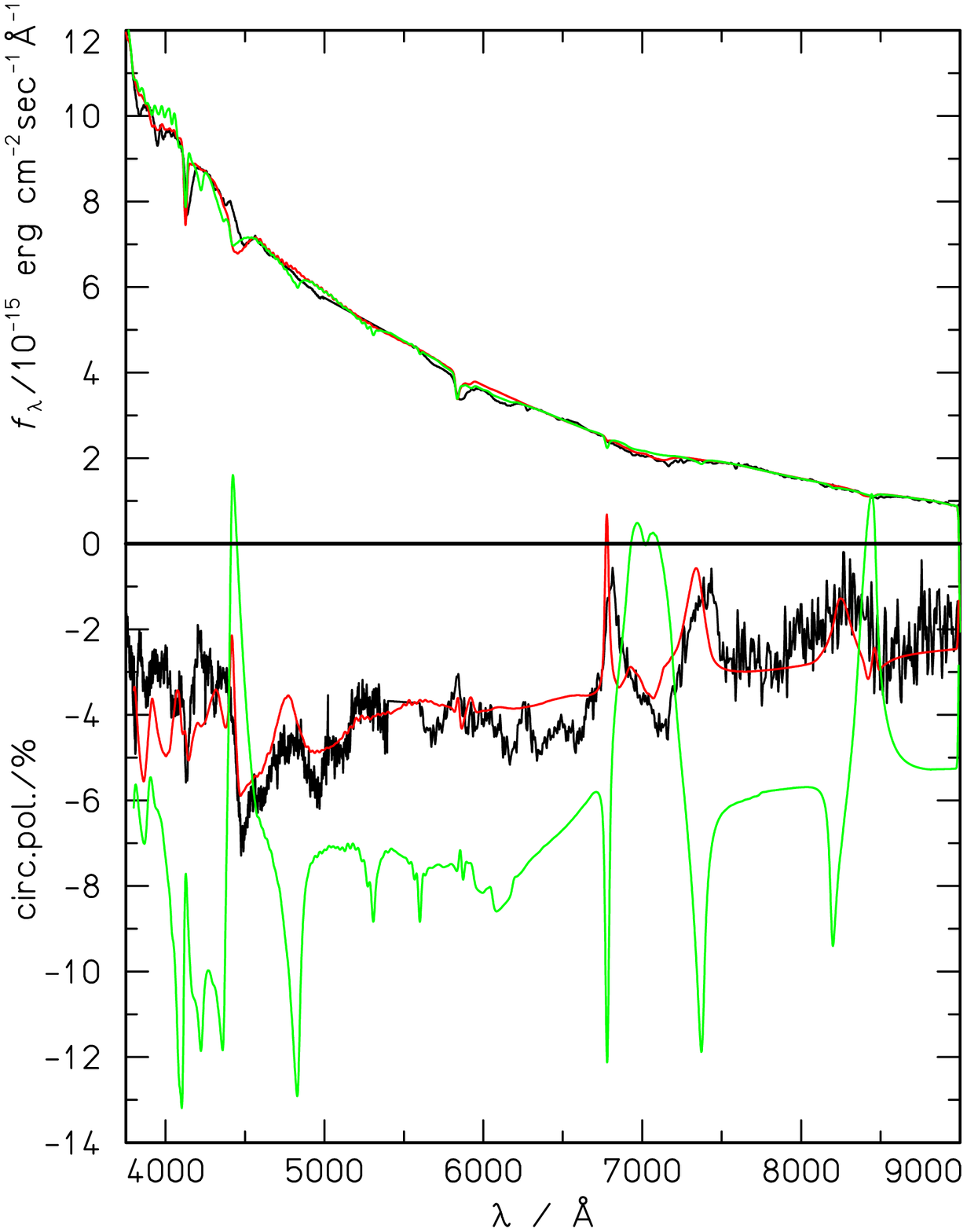}
\hspace{3cm}\epsfxsize=0.4\textwidth
\epsfbox{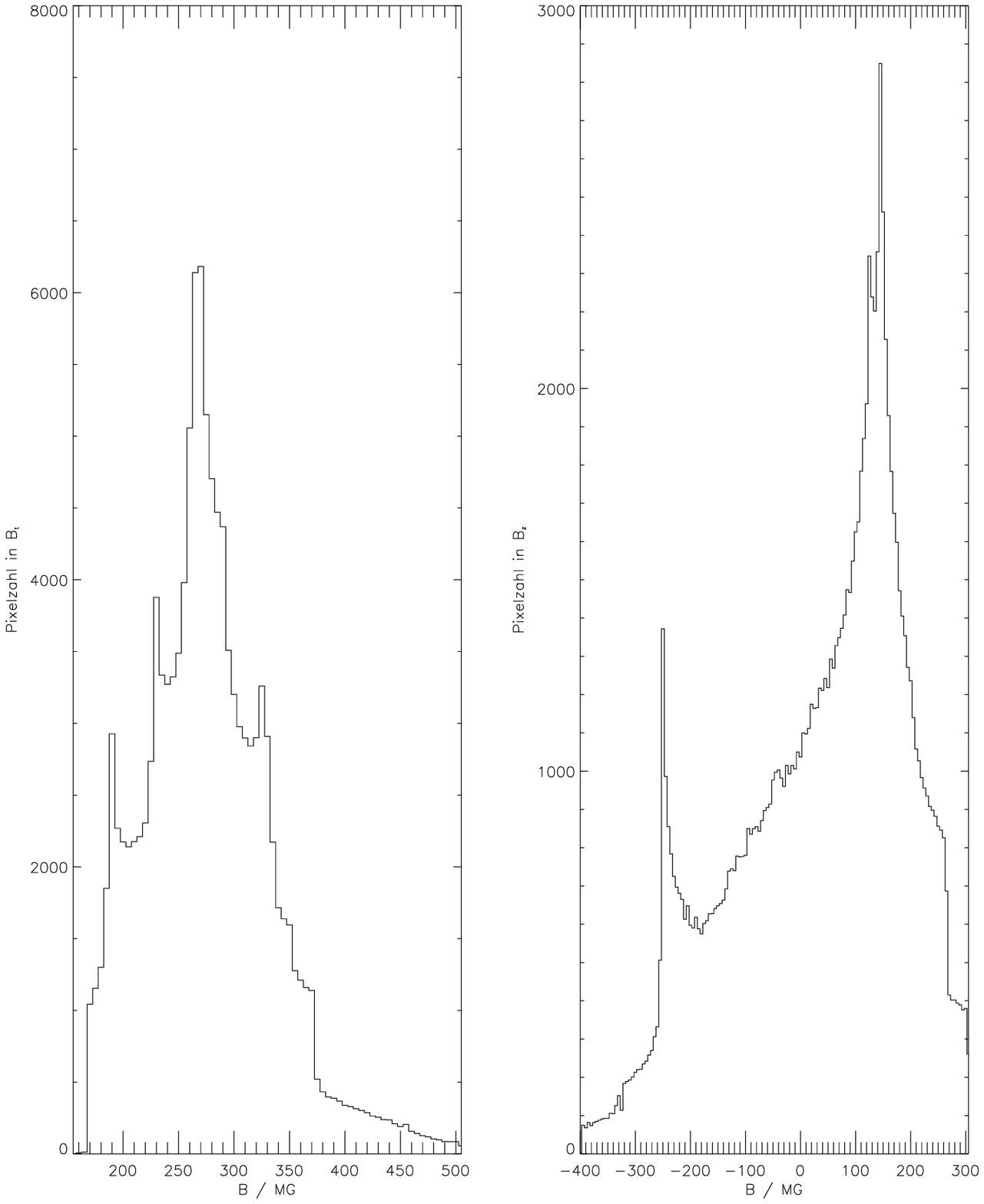}
\caption{Left: Synthetic spectra and circular polarization compared to the
observation of Grw~$+70^\circ 8247$ (black). The light grey curve 
corresponds to the best fit pure magnetic dipole model, while the best fit for an
expansion into spherical harmonics ($l \le 4$) is presented in  dark
grey. Right: Distribution of the absolute value of the magnetic field (left)
and of the component towards the observer (right) for the expansion
into spherical harmonics
}
\end{figure}

\begin{figure}[t]
\epsfxsize=0.322\textwidth
\epsfbox[51.6 42.0 350 652.5]{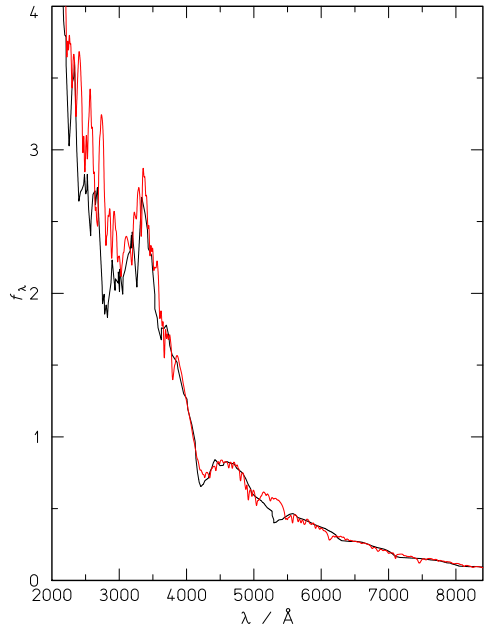}
\caption{Spectrum of GD\,229 (black) compared to a shifted dipole model
(grey)
}
\end{figure}

The next step was an 
 an expansion of the surface 
magnetic fields into spherical harmonics

\begin{displaymath}
\begin{array}{lll}
B_r=&-\sum_{l=1}^\infty \sum_{m=0}^l (l+1) (g_l^m \cos m\phi
+h_l^m \sin m \phi)\\ &\hfill  P_l^m(\cos\theta)\\
B_\theta=+ &\sum_{l=1}^\infty \sum_{m=0}^l  (g_l^m \cos m\phi
+ h_l^m \sin m \phi)\\ & \hfill dP_l^m(\cos\theta)/d\theta\\
B_\phi=&-\sum_{l=1}^\infty \sum_{m=0}^l  m (g_l^m \cos m\phi
+ h_l^m \sin m \phi)\\&\hfill dP_l^m(\cos\theta)/d\sin \theta
\end{array}
\end{displaymath}
with the associated Legendre polynomials $P_l^m$. The components 
are given in spherical coordinates $r, \theta,$ and $\phi$.
For this paper we limited ourselves to $l\le 4$,
i.e. to 24 free parameters; an additional parameter is the angle 
$i$ between the arbitrary magnetic axis and the observer. 

The best fit parameters were  

\begin{tabular}{cccccc}
$i=75.9^\circ$   \\
$g_{10}=183.00$  & 
$g_{11}=+0.71$ & $h_{11}=+0.36$ & \\
$g_{20}=-40.58$ &  
$g_{21}=+16.16$ & $h_{21}=-16.38$ & 
$g_{22}=+0.02$ & $h_{22}=+0.16$ \\
$g_{30}=+1.39$ & 
$g_{31}=-1.51$ & $h_{31}=+6.16$ & 
$g_{32}=-0.46$ & $h_{32}=-0.38$ \\ 
$g_{33}=-0.22$ & $h_{33}=+0.55$ \\
$g_{40}=+1.45$ & 
$g_{41}=-5.41$ & $h_{41}=+7.41$ & 
$g_{42}=-0.49$ & $h_{42}=-0.37$ \\ 
$g_{43}=+0.53$ & $h_{43}=+0.56$ & 
$g_{44}=+0.11$ & $h_{44}=-0.48$ \\
\end{tabular}

Fig.\,1 shows the improvement: While the fit to the flux spectrum remained
virtually unchanged, the main features of the wavelength dependent circular
polarization can now be well reproduced by the model.  The change
is mainly due to a different distribution of the magnetic field component
towards the observer. 
Due to the large number of free parameters it is clear that this
solution is not unique and that other  models could fit the flux
and polarization spectrum well. However, this is the first time that 
the circular polarization of the most famous magnetic white dwarf
Grw~$+70^\circ 8247$ could be explained by any model.

\section{Modeling of GD~229}
For a long time modeling of magnetic white dwarfs was limited to pure
hydrogen. However, as in the case of non-magnetic white dwarfs some
of the objects posses helium rich atmospheres. However, wavelengths for 
helium components were not available before \citeyear{JSBS98}
when \citeauthor{JSBS98}
could identify most of the absorption features in the spectrum of GD~229
with stationary line components. In the meantime the data sets have
been enlarged (see \citeauthor{JSB01} \shortcite{JSB01}). 

Radiative transfer calculations had to wait until  data for oscillator 
strengths were 
calculated by Schmelcher and Becken. Now a first application to GD~229
with consistent line data for helium in a magnetic field is presented
here. 

Fig.\,2 shows the best fit to the spectrum of GD~229 with an offset dipole
model ($i=30^\circ$, dipole strength 497\,MG, offset in dipole direction
$-0.25$ stellar radii, offsets in other directions $<0.004$ stellar radii)
and several features,
in particular the main feature at 4200\,\AA, are well reproduced. However,
several other parts of the spectrum lack resemblance with the observation.
There is also no agreement of the model prediction with circular polarization.
Therefore, the fitting is not satisfactory at the moment and has to be
continued.

\end{document}